\documentclass[11pt,dvips]{article}

\usepackage{epsfig}
\usepackage{graphicx,float}
\usepackage{array}
\usepackage{tabularx}
\usepackage{indentfirst}
\usepackage{amsmath,amssymb,bbold}
\usepackage{url}
\usepackage{rotating}
\usepackage{color}
\usepackage{hhline}
\usepackage{hyperref}
\usepackage{slashed}
\usepackage{tikz}
\usepackage{makecell}
\usepackage{pgfplots}
\usepackage{empheq}
\usepackage{physics}

\newcommand{\s}{\\ \vspace*{-3.5mm}}
\renewcommand{\thefootnote}{\alph{footnote}}

\begin{document}

% \begin{titlepage}
% \thispagestyle{empty}
% \setcounter{page}{0}

\begin{flushright}
%  aaa-nnnnn\\
%  hep--ph/yymmnnn\\[2mm]
% \today \\
\end{flushright}

\vskip 0.5cm

\begin{center}
{\Large \bf
 A Nondiagonal Pair of Majorana Particles
 at \boldmath{$e^+e^-$} Colliders}\\[1.0cm]
{Seong Youl Choi\footnote{sychoi@jbnu.ac.kr} and
 Jae Hoon Jeong\footnote{jaehoonjeong229@gmail.com}} \\[0.5cm]
{\it  Department of Physics and RIPC, Jeonbuk National University,
      Jeonju 54896, Korea}
\end{center}

\vskip 1.0cm

\begin{abstract}
\noindent
We perform a comprehensive and model-independent analysis for characterizing
the spin and dynamical structure of the production of a non-diagonal pair
of Majorana particles with different masses and arbitrary spins,
$e^-e^+\to X_2 X_1$, followed by a sequential two-body decay,
$X_2\to Z X_1$, of the heavier particle $X_2$ into a $Z$ gauge boson
and the lighter particle $X_1$ escaping undetected at high-energy
$e^+e^-$ colliders. Standard leptonic $Z$-boson decays,
$Z\to \ell^-\ell^+$ with $\ell=e$ or $\mu$, are employed for precisely
diagnosing the $Z$ polarization influenced by the production and decay
processes. Based on helicity formalism and Wick helicity rotation for
describing the correlated production-decay amplitudes and distributions,
we work out the implications on the amplitudes and distributions
of discrete CP symmetry and the Majorana condition that two particles are
their own anti-particles. For the sake of a concrete illustration, an example
of this type in the minimal supersymmetric Standard Model is investigated
in detail.
% \noindent PACS numbers: 12.60.-i, 12.60.Jv, 14.80.Ly
\end{abstract}

% \end{titlepage}

% \newpage

\vskip 0.5cm

\renewcommand{\thefootnote}{\fnsymbol{footnote}}

\section{Introduction}
\label{sec:introduction}

The Standard Model (SM)~\cite{Glashow:1961tr,Weinberg:1967tq,Salam:1968rm}
has been firmly confirmed as a self-consistent gauge theory with a weakly-coupled
sector for electroweak (EW) symmetry breaking with the discovery of a scalar
boson~\cite{Aad:2012tfa,Chatrchyan:2012ufa} and the ever-increasing confidence
of its compatibility with the SM Higgs boson~\cite{Aad:2019mbh,Sirunyan:2018koj}
at the CERN Large Hadron Collider (LHC). Nevertheless, we highly expect
new physics beyond the SM (BSM) to be revealed at the TeV scale (Terascale),
motivated by tiny but non-vanishing neutrino masses~\cite{GonzalezGarcia:2007ib},
matter dominance in our Universe~\cite{Morrissey:2012db,Buchmuller:2005eh},
dark matter~\cite{Jungman:1995df,Bertone:2004pz,Dine:2003ax} and
inflation~\cite{Lyth:1998xn}, etc. Conceptually, the naturalness
issue~\cite{Gildener:1976ih,Weinberg:1975gm,Susskind:1978ms}
has been the prime argument for the realization of new BSM physics
at the weak scale of 246 GeV.\s

To much puzzlement, except for a SM-like Higgs boson, no new BSM
particles have been so far observed in the LHC experiments around
the Terascale threshold. One plausible scenario for the LHC null search
results is that all the strongly-interacting colored BSM particles are
too heavy to be directly produced at the LHC and the electroweak (EW)
BSM particles, although kinematically accessible, may not lead to tractable
signals due to rather small production rate, uncharacteristic signature
and/or large SM backgrounds at hadron colliders. On the other hand,
the future high-energy $e^+e^-$ colliders  would be capable of discovering
and diagnosing some new EW particles, as long as kinematically accessible,
because of well-constrained event topology and very clean experimental
environment.\s

In the present work, we study such a challenging but plausible scenario
at an $e^+e^-$ collider that two neutral BSM particles are kinematically
accessible only with the combination of the diagonal pair of the lighter
particles and the non-diagonal pair of two particles, while the diagonal
pair of the heavier particles is kinematically inaccessible, see for example Ref.~\cite{Choi:2001ww}. Particularly, two neutral particles are assumed
to be their own antiparticles but their spins are arbitrary. 
Such Majorana particles~\footnote{Usually
the term Majorana has been used for fermions with half-integer spin but
it will be employed for real bosons with integer spin as well.} are unavoidable
in supersymmetric theories, guaranteeing that every known bosonic particle
has a heavier fermionic partner and vice versa for each known fermion,
and they are predicted also by various grand unified theories and
extra-dimensional models and even in solid-state physics~\cite{Elliott:2014iha}.\s

Referring to Refs.~\cite{Choi:2015afa,Choi:2015zka} as a few previous works
for the processes of diagonal pair production, we focus on the analysis of
the combined process of production of a non-diagonal pair of Majorana
particles with different masses and arbitrary spins, followed by a sequential
decay chain of two-body decays as
\begin{eqnarray}
   e^-\, +\, e^+\, &\to& \, X_2\, + \, X_1 \nonumber \\[1.5mm]
                   & & \ \ \rotatebox[origin=c]{180}{\Large $\Lsh$}
                       \ \  Z\, + \, X_1 \nonumber\\[1.5mm]
                   & & \ \ \qquad \rotatebox[origin=c]{180}{\Large $\Lsh$}
                       \ \  \ell^-\, +\, \ell^+\,,
\label{eq:combined_eex2x1_x2zx1}
\end{eqnarray}
%
% \rotatebox[origin=c]{180}{\huge $\Lsh$}
where the mass splitting $\Delta m= m_2-m_1$ of the particles $X_2$ and $X_1$
is larger than the $Z$ boson mass $m_Z$, i.e. $m_2-m_1> m_Z$, and
the charged lepton $\ell$ is taken to be $e$ or $\mu$, allowing for
the full reconstruction of the $Z$-boson momentum with great
precision.\footnote{If $\Delta m < m_Z$, the particle $X_2$ decays directly
into  $X_1\ell^-\ell^+$ through several channels. These three-body decay
processes are closely related to the production process $e^-e^+\to X_2 X_1$,
especially for $\ell=e$, from topological point of view. A detailed
analysis of these combined production-decay process involving a few
sophisticated conceptual issues will be reported elsewhere.}\s

The neutral particle $X_1$ is assumed to be stable and so it escapes undetected
with no tractable signals as the lightest neutral supersymmetric particle (LSP)
in the minimal supersymmetric SM (MSSM) with $R$ parity. Consequently, the
combined process has a distinct $V$-shape signature of a charged-lepton pair
of which the four-momentum is balanced due to energy-momentum conservation
with the missing four-momentum carried away by two invisible $X_1$ particles
\begin{eqnarray}
e^-e^+\ \ \rightarrow\ \ \ell^-\ell^+\, +\, \not\!\!{E}\,,
\end{eqnarray}
with $\not\!\!{E}$ denoting the invisible part and with the constraint 
$m_{\ell\ell}=m_Z$ for the invariant mass of two final leptons, signalling 
the presence of an intermediate on-shell $Z$ boson. \s

For a non-zero $X_2$ spin $j_2$, the $X_2$ particle is produced generally in a
polarized state in the process (\ref{eq:combined_eex2x1_x2zx1}),
especially, if the interactions are parity-violating. The information on
the polarization of the spin-1 gauge boson $Z$ can be extracted through the 
angular distributions in its well-established leptonic decays, $Z\to \ell^-\ell^+$,
with $\ell=e$ or $\mu$. However,
the $Z$ momentum direction in the $X_2$ rest frame ($X_2$RF) which is the most
convenient for describing the decays analytically  is not identical
to that in the $e^+e^-$ CM frame ($ee$CM) directly reconstructed experimentally.
A proper Wick helicity rotation \cite{Leader:2001gr,Choi:2018sqc,Choi:2019aig}
needs to be incorporated for linking the $Z$ polarization state with respect to the
$Z$ momentum direction in the $ee$CM to that with respect to the $Z$ momentum
direction in the $X_2$RF. \s

The prime goal of the present work is to derive the correlated production-decay
(polar-)angular correlations in a transparent and compact way exploiting
the helicity formalism and an Wick helicity rotation for probing the
spins and dynamical properties of two Majorana particles in a general setting.\s

The paper is organized as follows. In Section~\ref{sec:helicity_amplitudes},
we present the complete amplitudes of the production process and two sequential
two-body decays in a compact and general form and analyze the implications
of the discrete CP symmetry and the Majorana condition on the amplitudes
and production-decay correlations.  Section~\ref{sec:correlated_angular_distributions}
is devoted to a systematic derivation of all the angular correlations and
a detailed analysis of the fully-reconstructible polar-angle correlations.
In Section~\ref{sec:two_specific_examples} those model-independent theoretical 
results are demonstrated with one specific example of a non-diagonal pair of
neutralinos~\cite{Ellis:1983er,Bilenky:1985wu,Bilenky:1986nd,MoortgatPick:2002iq,
Khristova:1987xq,Balantekin:2018ukw}, which are Majorana fermions in the MSSM.
Finally some conclusions are given in Section~\ref{sec:conclusions}. \s

\section{Production and Decay Amplitudes}
\label{sec:helicity_amplitudes}

In this section, firstly we present in a compact and transparent form the complete
helicity amplitudes for the production of a nondiagonal pair of Majorana particles
and for the two sequential two-body decays shown in 
Eq.$\,$(\ref{eq:combined_eex2x1_x2zx1}), of which the kinematical configuration is 
depicted in detail
in Figure~\ref{fig:kinematical_configuration_combined_production_decay}.
A proper Wick helicity rotation and an azimuthal-angle adjustment are performed
for linking the decay helicity amplitudes in the $X_2$RF and $ee$CM. Secondly,
we remark on the general constraints on the amplitudes by CP invariance and
by the Majorana condition.  \s

\subsection{Helicity amplitudes}
\label{subsec:helicity_amplitudes}

We adopt the helicity formalism~\cite{Leader:2001gr,Wick:1962zz} for deriving
the helicity amplitudes of the production process for a nondiagonal pair $X_2 X_1$
of Majorana particles
\begin{eqnarray}
    e^-(k,\sigma)\, +\, e^+(\bar{k},\bar{\sigma})
\ \ \rightarrow \ \
    X_2(p_2,\lambda_2)\, +\, X_1(p_1,\lambda_1)\,,
\end{eqnarray}
in the $e^-e^+$ center-of-mass (CM) frame ($ee$CM), and
those of the two-body decay of the Majorana particle $X_2$ of mass $m_2$ and
spin $j_2$ into an on-shell spin-1 $Z$ boson of mass $m_Z$ and a Majorana
particle $X_1$ of mass $m_1$ and spin $j_1$
\begin{eqnarray}
X_2(p^\prime_2,\lambda_2)
   \ \ \rightarrow\ \
Z(q_Z,\lambda_Z)\, +\, X_1(q_1,\sigma_1)\,,
\end{eqnarray}
in the $X_2$RF (See Refs.$\,$\cite{Choi:2018sqc,Choi:2003fs} for the
neutralino two-body decay in the MSSM) and for the $Z$ two-body leptonic
decay
\begin{eqnarray}
Z(q^\prime_Z,\lambda^\prime_Z)
   \ \ \rightarrow\ \
\ell^-(k_-,\tau_-)\, + \ell^+(k_+,\tau_+)\,,
\end{eqnarray}
in the $Z$ rest frame ($Z$RF). The four-momentum and helicity of each particle
are shown in parenthesis with each primed four-momentum referring to the
four-momentum in the rest frame of its corresponding particle, $X_2$ or $Z$.
One crucial point to be ensured in calculating the amplitudes of the correlated
production-decay process is that the $Z$-boson polarization state in 
the $X_2$RF is in general different from that in the $Z$RF directly reconstructible
in the $ee$CM. \s 

% \vskip 1.cm

%
\begin{figure}[tbh]
\begin{center}
\includegraphics[width=10.cm, height=10cm]{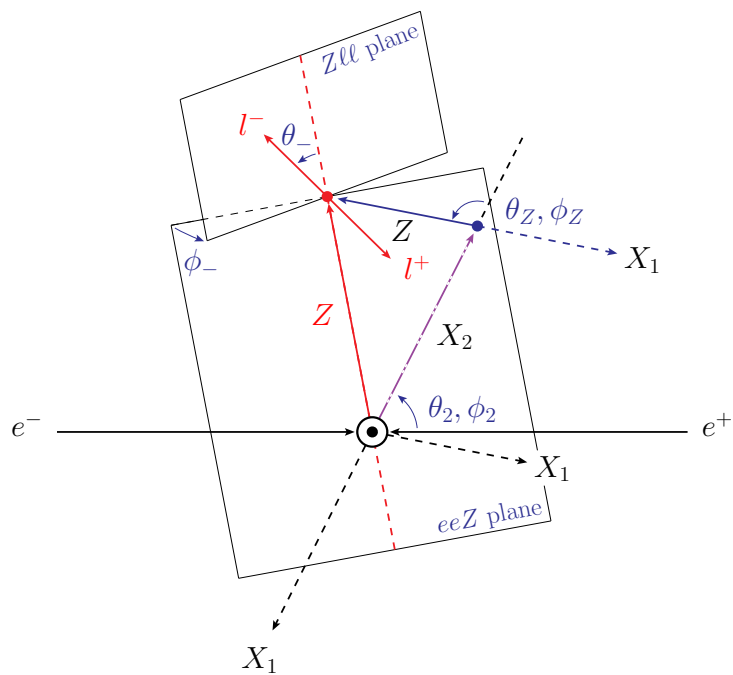}
\caption{\it A diagrammatic description of kinematical configurations of the combined
             production-decay process (\ref{eq:combined_eex2x1_x2zx1}). The $X_2$ polar
             and azimuthal angles, $\theta_2$ and $\phi_2$, are defined with respect to
             the $e^-$ momentum direction and a properly chosen $x$-axis. 
             The $Z$ polar and azimuthal angles, $\theta_Z$
             and $\phi_Z$, are defined in the $X_2$ rest frame boosted back along
             the $X_2$ momentum direction from the $ee$CM and the $\ell^-$ polar and
             azimuthal angles, $\theta_-$ and $\phi_-$, are
             defined in the $Z$ rest frame boosted back along the $Z$ momentum
             direction in the $ee$CM, respectively. The azimuthal angle $\phi_-$ denotes
             the relative angle between the $eeZ$ plane and the $Z\ell\ell$ plane
             in the $ee$CM.
}
\label{fig:kinematical_configuration_combined_production_decay}
\end{center}
\end{figure}

Explicitly, in the kinematical configuration depicted
in Figure~\ref{fig:kinematical_configuration_combined_production_decay},
the helicity amplitude of the production process $e^-e^+\to X_2 X_1$ can be
written as
\begin{eqnarray}
 {\cal M}^{e^-e^+\to X_2 X_1}_{\sigma,\bar{\sigma};\lambda_2,\lambda_1}
 (\theta_2,\phi_2)
= {\cal P}_{\sigma,\bar{\sigma};\lambda_2,\lambda_1}(\cos\theta_2)\,\,
  d^J_{\sigma-\bar{\sigma}, \lambda_2-\lambda_1}(\theta_2)\,
  e^{\imath (\sigma-\bar{\sigma})\phi_2}\,,
\label{eq:production_helicity_amplitudes}
\end{eqnarray}
where $J={\rm max}(|\sigma-\bar{\sigma}|, |\lambda_2-\lambda_1|)$, and the angles
$\theta_2$ and $\phi_2$ denote the scattering polar and azimuthal angles of
the $X_2$ with respect to the $e^-$ momentum direction and a fixed $x$-axis, of which
the direction may be fixed for transverse $e^-$ or $e^+$ beam polarizations,
in the $ee$CM. Finally the polar-angle dependent function
$d^J_{\sigma-\bar{\sigma},\lambda_2-\lambda_1}(\theta_2)$
is the Wigner $d$ function in the convention of Rose~\cite{merose2011}.\s

The general theoretical analysis of the $Z$ polarization in the two-body
decay $X_2\to Z X_1$ is the most transparent analytically in the $X_2$RF. 
The decay helicity amplitude can be decomposed in terms of the decay polar 
and azimuthal angles, $\theta_Z$ and $\phi_Z$, for the momentum direction 
of the $Z$ boson in the $X_2$RF
\begin{eqnarray}
  {\cal M}^{X_2\to Z X_1}_{\lambda_2;\lambda_Z,\sigma_1}(\theta_Z,\phi_Z)
= {\cal C}_{\lambda_Z,\sigma_1}\,
  d^{j_2}_{\lambda_2,\lambda_Z-\sigma_1}(\theta_Z)\,
  e^{i\lambda_2\phi_Z}
  \quad\mbox{with}\quad
  |\lambda_Z-\sigma_1|\leq j_2\,,
\label{eq:x2_zx1_helicity_amplitude}
\end{eqnarray}
where the azimuthal angle $\phi_Z$ is defined with respect to the plane formed by
the $e^-$ and $X_2$ momenta in the $ee$CM. Because the $Z$-momentum direction
in the $X_2$RF is different from that in the $ee$CM, the helicity amplitudes
in Eq.$\,$(\ref{eq:x2_zx1_helicity_amplitude}) need to be transformed
by a proper Wick helicity rotation \cite{Leader:2001gr,Choi:2018sqc,Choi:2019aig}
for connecting the $Z$ helicity state in the $X_2$RF to that in the $ee$CM 
with a so-called Wick helicity rotation angle $\omega_{_Z}$ satisfying
\begin{eqnarray}
    \cos\omega_{_Z}
&=& \frac{\beta_Z+\beta_2\cos\theta_Z}{
          \sqrt{(1+\beta_2\beta_Z\cos\theta_Z)^2-(1-\beta^2_2)(1-\beta^2_Z)}}\,,\\
    \sin\omega_{_Z}
&=& \frac{\sqrt{1-\beta^2_Z}\, \beta_2\sin\theta_Z}{
          \sqrt{(1+\beta_2\beta_Z\cos\theta_Z)^2-(1-\beta^2_2)(1-\beta^2_Z)}}\,,
\label{eq:cos_sin_wick_helicity_rotation_angle}
\end{eqnarray}
where $\beta_2$ and $\beta_Z$ are the $X_2$ speed in the $ee$CM and
the $Z$ speed in the $X_2$RF, which are unambiguously determined
in terms of the $e^-e^+$ CM energy $\sqrt{s}$ and the $X_{1,2}$ and
$Z$-boson masses. The resulting decay helicity amplitude directly coupled
with the $Z$-boson decay helicity amplitude reads~\footnote{We do not include
another Wick helicity rotation connecting the $X_1$ helicity states in
the $ee$CM and in the $X_2$RF because its effects on any distributions are
washed away completely with summing over the $X_1$ helicities, naturally
taken for the invisible $X_1$ particle.}
\begin{eqnarray}
  {\cal A}_{\lambda_2;\lambda^\prime_Z,\sigma_1}(\theta_Z,\phi_Z)
= \sum_{\lambda_Z=\pm 1, 0}\, d^1_{\lambda^\prime_Z,\lambda_Z}(\omega_{_Z})\,\,
  {\cal M}^{X_2\to Z X_1}_{\lambda_2;\lambda_Z,\sigma_1}(\theta_Z,\phi_Z)\,,
\label{eq:wick_helicity_rotated_x2_zx1_decay_helicity_mplitude}
\end{eqnarray}
It is important to note that the Wick helicity rotation angle $\omega_{_Z}$ along
with the polar angle $\theta_Z$ is determined event by event, although the azimuthal
angle $\phi_Z$ defined with respect to the $e^- X_2$ plane in the $ee$CM cannot
be reconstructed due to the invisible $X_1$. \s

Among various decay channels of the $Z$ boson, the leptonic $Z$-boson decays
$Z\to\ell^-\ell^+$, especially with $\ell=e$ and $\mu$, provide a very clean
and powerful means for reconstructing the $Z$-boson rest frame, independently
of its production mechanisms, and for extracting the information on $Z$
polarization. The helicity amplitude of the leptonic $Z$-boson decay can
be written as
\begin{eqnarray}
  {\cal M}^{Z\to \ell^-\ell^+}_{\lambda^\prime_Z;\,\sigma_-,\sigma_+}(\theta_-,\phi_-)
= {\cal Z}_{\sigma_-,\sigma_+}\,
  d^{1}_{\lambda^\prime_Z,\sigma_--\sigma_+}(\theta_-)\,
  e^{i\lambda^\prime_Z \phi_-}\,,
\label{eq:z_ll_helicity_amplitude}
\end{eqnarray}
in terms of the polar and azimuthal angles, $\theta_-$ and $\phi_-$, in the $Z$RF
with the azimuthal angle defined with respect to the plane formed by the
$e^-$ and $Z$ momenta in the $ee$CM, which are determined fully with great
precision. In terms of the normalized vector and axial-vector couplings
$v_\ell=\sin^2\theta_W-1/4$ and $a_\ell=1/4$ with the weak mixing
angle $\theta_W$, the reduced helicity amplitude ${\cal Z}_{\sigma_-,\sigma_+}$
in Eq.$\,$(\ref{eq:z_ll_helicity_amplitude}) is given by
\begin{eqnarray}
  Z_{\sigma_-,\sigma_+}
= -\imath \sqrt{2}\, g_{_Z}\, m_Z \,
  \left(v_\ell+\Delta\sigma\, a_\ell\right)\,\, \delta_{\sigma_-,-\sigma_+}\,,
\end{eqnarray}
with $\Delta\sigma=\sigma_- - \sigma_+=\pm 1$ and $g_{_Z}=e/c_W s_W$
in terms of the positron electric charge $e$ and the abbreviations, 
$c_W=\cos\theta_W$ and $s_W=\sin\theta_W$, when the charged lepton mass is ignored.
However, it is necessary to adjust the azimuthal-angle phase factor
of the helicity amplitude in Eq.$\,$(\ref{eq:z_ll_helicity_amplitude})
by an azimuthal angle $\gamma_Z$ for compensating the mismatch between
the $ZX_2X_1$ plane and the $eeZ$ plane in the $ee$CM, leading to the $Z$-boson
decay amplitude with an adjusted phase factor as
\begin{eqnarray}
   {\cal B}_{\lambda^\prime_Z;\sigma_-\sigma_+}(\theta_-,\phi^\prime_-)
= {\cal Z}_{\sigma_- \sigma_+}\,
    d^{1}_{\lambda^\prime_Z,\sigma_--\sigma_+}(\theta_-)\,\,
    e^{i\lambda^\prime_Z \phi^\prime}\,,
\label{eq:adjusted_z_ll_decay_amplitude}
\end{eqnarray}
with the newly-defined azimuthal angle $\phi^\prime=\phi-\gamma_{_Z}$. The angle
$\gamma_{_Z}$ satisfies
\begin{eqnarray}
    \cos\gamma_{_Z}
&=& \frac{\cos\theta_2\sin\theta^\prime_Z+\sin\theta_2\cos\theta^\prime_Z\cos\phi_Z}{
     \sqrt{1-(\cos\theta_2\cos\theta^\prime_Z
             -\sin\theta_2\sin\theta^\prime_Z\cos\phi_Z)^2}}\,,\\
    \sin\gamma_{_Z}
&=& \frac{\sin\theta_2\sin\phi_Z}{
     \sqrt{1-(\cos\theta_2\cos\theta^\prime_Z
             -\sin\theta_2\sin\theta^\prime_Z\cos\phi_Z)^2}}\,,
\label{eq:cos_sin_gamma_Z}
\end{eqnarray}
where the angle $\theta^\prime_Z$ is the $Z$-boson polar angle with respect to the
$X_2$ momentum direction in the $ee$CM, which can be determined event by event
through the relations
\begin{eqnarray}
    \tan\theta^\prime_Z
= \frac{\beta_Z\sin\theta_Z}{\gamma_2(\beta_2+\beta_Z\cos\theta_Z)}
 \quad \mbox{and}\quad
     E'_Z
= m_Z\,\gamma_2\gamma_Z (1+\beta_2\beta_Z\cos\theta_Z)\,,
\label{eq:gamma_theta_Z_E_Z_prime_relation}
\end{eqnarray}
with $\gamma_2=(s+m^2_2-m^2_1)/2m_2\sqrt{s}$, $\beta_2=\sqrt{1-1/\gamma^2_2}$,
$\gamma_Z=(m^2_2-m^2_1+m^2_Z)/2m_2m_Z$ and $\beta_Z=\sqrt{1-1/\gamma^2_Z}$,
and with the polar angle $\theta_Z$ determined by measuring the $Z$-boson energy $E'_Z$
in the $ee$CM directly, as can be checked with the right expression in
Eq.$\,$(\ref{eq:gamma_theta_Z_E_Z_prime_relation}). However, the polar angle
$\theta_2$ and the azimuthal angle $\phi_Z$ cannot be directly measured
event by event because of two invisible $X_1$ particles in the combined
production-decay process (\ref{eq:combined_eex2x1_x2zx1}).\s

Combining the production helicity amplitude in Eq.$\,$(\ref{eq:production_helicity_amplitudes})
and two decay helicity amplitudes in Eqs.$\,$(\ref{eq:wick_helicity_rotated_x2_zx1_decay_helicity_mplitude})
and (\ref{eq:adjusted_z_ll_decay_amplitude}) adjusted by an Wick helicity
rotation and an azimuthal rotation, we obtain the fully-correlated
production-decay helicity amplitude as
{\small
\begin{eqnarray}
  {\cal M}_{\sigma,\bar{\sigma};\sigma_-,\sigma_+;\lambda_1,\sigma_1}
= D_2(p^2_2)D_{_Z}(q^2_Z) \sum_{\lambda_2,\lambda^\prime_Z}
  {\cal M}_{\sigma,\bar{\sigma};\lambda_2,\lambda_1}(\theta_2,\phi_2)\,
  {\cal A}_{\lambda_2;\lambda^\prime_Z,\sigma_1}(\theta_Z,\phi_Z)\,
  {\cal B}_{\lambda^\prime_Z;\sigma_-,\sigma_+}(\theta_-,\phi^\prime_-)\,,
\label{eq:fully_combined_amplitudes}
\end{eqnarray}
}
\vskip -0.3cm
\noindent
with the adjusted azimuthal angle $\phi^\prime_- =\phi_- - \gamma_{_Z}$ and
the $X_2$ and $Z$ Breit-Wigner propagators,
$D_2=1/(p^2_2-m^2_2+i m_2\Gamma_2)$ and
$D_Z=1/(q^2_Z-m^2_Z+im_Z\Gamma_Z)$.\s

\subsection{CP symmetry and Majorana Condition}
\label{subsec:cp_symmetry_majorana_condition}

Before going into a detailed description of the angular correlations
in Section~\ref{sec:correlated_angular_distributions}, we study some
general restrictions on the helicity amplitudes
due to CP invariance and the Majorana condition that each of the neutral
particles $X_2$ and $X_1$ is its own antiparticle, respectively.\s

Even in transitions involving weak interactions, the production and decay
processes observe CP symmetry to a great extent while often violating
P and C symmetries. So we discuss the consequences of the CP symmetry
among discrete spacetime symmetries in the production and decay helicity
amplitudes. For the production and decay processes involving two Majorana
particles $X_2$ and $X_1$, CP invariance leads to the following relations
\begin{eqnarray}
    {\cal P}_{\sigma,\bar{\sigma}; \lambda_2,\lambda_1} (\cos\theta_2)
&=& \eta^P_{_{\rm CP}}\, (-1)^J\,\,
    {\cal P}_{-\bar{\sigma},-\sigma; -\lambda_2,-\lambda_1}(-\cos\theta_2)\,,
\label{eq:production_cp_relation}\\
    {\cal C}_{\lambda_Z,\sigma_1}
&=& \eta^C_{_{\rm CP}}\,\, {\cal C}_{-\lambda_Z,-\sigma_1}\,,
\label{eq:x2_z_x1_cp_relation}
\end{eqnarray}
with the appropriate helicity-independent CP parities, $\eta^P_{_{\rm CP}}$
and $\eta^C_{_{\rm CP}}$, consisting of intrinsic parties and particle spins.
Note that these CP tests do not assume the absence of absorptive parts and
rescattering effects at all.\s

Together with CPT invariance valid in the absence of absorptive parts 
and/or rescattering effects, the Majorana condition that both of the two neutral
particles $X_2$ and $X_1$ are their own antiparticles leads to the
relations for the production and decay helicity amplitudes:
\begin{eqnarray}
    {\cal P}_{\sigma,\bar{\sigma}; \lambda_2,\lambda_1} (\cos\theta_2)
&=& \eta^P_{_{\rm M}}\, (-1)^J\,\,
    {\cal P}^*_{-\bar{\sigma},-\sigma; -\lambda_2,-\lambda_1}(-\cos\theta_2)\,,
\label{eq:production_majorana_condition}\\
    {\cal C}_{\lambda_Z,\sigma_1}
&=& \eta^C_{_{\rm M}}\,\, {\cal C}^*_{-\lambda_Z,-\sigma_1}\,,
\label{eq:x2_z_x1_majorana_condition}
\end{eqnarray}
where the parity factors $\eta^{P}_{\rm M}$ and $\eta^{C}_{\rm M}$ are
dependent on the intrinsic CPT parities and spins but independent of
helicities.\s

\section{Correlated Angular Distributions}
\label{sec:correlated_angular_distributions}

The fully-correlated production-decay amplitudes
in Eq.($\,$\ref{eq:fully_combined_amplitudes}) allow us to probe all
the polarization phenomena with which the spins and interaction structures
of the production and decay processes can be determined. In this Section,
we derive all the analytic expressions for the correlated angular distributions,
which consist of three helicity-dependent parts.\s

The first process under attack is the production of a non-diagonal
pair of Majorana particles $e^-e^+\to X_2 X_1$. Summing over the helicities
of the invisible $X_1$ and incorporating the electron and positron $2\times 2$
polarization density matrices, $\rho^-$ and $\rho^+$,
we can write the helicity-dependent differential cross section in the form
\begin{eqnarray}
 \frac{d\sigma}{d\Omega_2}(\lambda_2,\lambda^\prime_2)
= \frac{\kappa_{21}}{64 \pi^2 s}\,\,
\Sigma^{\lambda_2}_{\lambda^\prime_2}\,,
\label{eq:helicity_dependent_production_cross_section}
\end{eqnarray}
where $d\Omega_2=d\cos\theta_2\, d\phi_2$, $\mu_{1,2}=m_{1,2}/\sqrt{s}$ and
$\kappa_{21}=  \lambda^{1/2}(1,\mu^2_2,\mu^2_1)$ with
the K\"{a}ll\'{e}n kinematical function $\lambda(1,x,y) = [1-(x+y)^2][1-(x-y)^2]$.
The production tensor $\Sigma$ in
Eq.$\,$(\ref{eq:helicity_dependent_production_cross_section}) reads
\begin{eqnarray}
  \Sigma^{\lambda_2}_{\lambda^\prime_2}
= \sum_{\sigma,\sigma^\prime}\,
  \sum_{\bar{\sigma},\bar{\sigma}^\prime}\,
  \sum_{\lambda_1}\,
  \rho^-_{\sigma,\sigma^\prime}\,
  \rho^+_{\bar{\sigma},\bar{\sigma}^\prime}\,\,
  {\cal M}_{\sigma,\bar{\sigma};\lambda_2,\lambda_1}
  {\cal M}^*_{\sigma^\prime,\bar{\sigma}^\prime; \lambda^\prime_2,\lambda_1}\,,
\label{eq:production_tensor}
\end{eqnarray}
with the implied summation over repeated indices $(\sigma,\sigma^\prime, \bar{\sigma},
\bar{\sigma^\prime})=\pm 1/2 =\pm$ and $\lambda_1=-j_1,\cdots, j_1$.
The $(2j_2+1)\times (2j_2+1)$ polarization density matrix of
the produced $X_2$ is given by
\begin{eqnarray}
  \rho^{X_2}_{\lambda_2,\lambda^\prime_2}
= \Sigma^{\lambda_2}_{\lambda^\prime_2}/\Sigma^{\sigma_2}_{\sigma_2}\,,
\label{eq:x2_production_density_matrix}
\end{eqnarray}
with the implied summation over the repeated $X_2$ helicity index
$\sigma_2=-j_2,\cdots,j_2$.\s

If only the longitudinal polarizations
\footnote{Transversely-polarized beams are not considered
in the present work because their
effects will be washed out after integrating the distributions over
the production azimuthal angle.} of the $e^-$ and $e^+$ beams and
the electron chirality conservation related to the tiny electron
mass~\cite{Hikasa:1985qi} is imposed on the electron-positron current,
the combined $e^-e^+$ polarization tensor is simplified as
\begin{eqnarray}
\rho^-_{\sigma,\sigma^\prime}\,
  \rho^+_{\bar{\sigma},\bar{\sigma}^\prime}
= \delta_{\sigma,\sigma^\prime}\,
  \delta_{\sigma,-\bar{\sigma}}\,
  \delta_{\sigma^\prime,-\bar{\sigma}^\prime}\,\,
  \frac{1}{4}\, \left[\, 1-P_L \bar{P}_L+\sigma\, (P_L-\bar{P}_L)\,\right]\,,
\label{eq:ee_longitudinal_polarization_density_matrix}
\end{eqnarray}
with the degrees $P_L$ and $\bar{P}_L$ of electron and positron longitudinal
polarizations, respectively. Because of the Majorana condition
(\ref{eq:production_majorana_condition}), the polar-angle distribution set by
the trace of the production tensor is forward-backward (FB) symmetric but the
P-odd $X_2$ polarization components defined by the differences
$\rho^{X_2}_{\lambda_2,\lambda_2}-\rho^{X_2}_{-\lambda_2,-\lambda_2}$
with $\lambda_2=-j_2, \cdots , j_2$ is FB antisymmetric.\s

In the narrow-width approximation, the produced $X_2$ particle decays
on-shell with good approximation. As pointed out before, it is necessary to
include an Wick helicity rotation and an azimuthal-angle adjustment for
calculating the helicity amplitude of the sequential decay chain of two
2-body decays $X_2\to Z X_1$ and $Z\to\ell^-\ell^+$. The correlated
decay distribution including the matrix in Eq.$\,$(\ref{eq:x2_production_density_matrix})
encoding $X_2$ polarization is given by
\begin{eqnarray}
  \frac{d\Gamma}{d\Omega_Z d\Omega_-}
 = \frac{3\kappa_Z}{256\pi^3 m_2}\,
   {\rm B}(Z\to\ell^-\ell^+)\,
   \sum_{\lambda_2,\sigma_2}
   \sum_{\lambda^\prime_Z,\sigma^\prime_Z}\,
   \sum_{\sigma_1}
    \rho^{X_2}_{\lambda_2,\sigma_2}\,
   \left[ {\cal A}_{\lambda_2;\lambda^\prime_Z,\sigma_1}\,
   {\cal A}^*_{\sigma_2;\sigma^\prime_Z,\sigma_1}\right]\,
    \rho^Z_{\lambda^\prime_Z,\sigma^\prime_Z}\,,
\label{eq:fully_correlated_distributions}
\end{eqnarray}
where $\kappa_Z=\lambda^{1/2}(1,m^2_1/m^2_2,m^2_Z/m^2_2)$,
$d\Omega_Z = d\cos\theta_Z d\phi_Z$ and $d\Omega_-=d\cos\theta_-
d\phi_-$ and the summation over all repeated helicity indices is
taken. With the known $Z\ell\ell$ couplings in the SM, the normalized
$3\times 3$ $Z$-boson decay density matrix $\rho^Z$ is given in terms
of an asymmetry parameter $A_\ell=2 v_\ell a_\ell/(v^2_\ell + a^2_\ell)$
by
{\small
\begin{eqnarray}
 \rho^Z_{\lambda^\prime_Z,\sigma^\prime_Z}(\theta_-,\phi^\prime_-)
 \, =\, \frac{1}{4}
    \left(\begin{array}{ccc}
       1 + c^2_- + 2 A_\ell\, c_-
     & \sqrt{2}\,(A_\ell + c_-)\, s_-\, e^{i\phi^\prime_-}
     & s^2_- \, e^{2i\phi^\prime_-}\\[2mm]
       \sqrt{2}\,(A_\ell + c_-)\, s_-\, e^{-i\phi^\prime_-}
     & 2 s^2_-
     & \sqrt{2}\,(A_\ell - c_-)\, s_-\, e^{i\phi^\prime_-}\\[2mm]
        s^2_- \, e^{-2i\phi^\prime_-}
     & \sqrt{2}\,(A_\ell - c_-)\, s_-\, e^{-i\phi^\prime_-}
     & 1 + c^2_- - 2 A_\ell\, c_-
        \end{array} \right)\,,
\label{eq:z_ll_decay_density_matrix}
\end{eqnarray}
}
\vskip -0.3cm
\noindent
in the $(+1,0,-1)$ helicity basis of the $Z$ boson with the abbreviations,
$c_-=\cos\theta_-$ and $s_-=\sin\theta_-$, and with the adjusted azimuthal
angle $\phi^\prime_-=\phi_- - \gamma_Z$. We emphasize once more
that the azimuthal angle $\gamma_Z$ depends on the $X_2$ polar angle
and $Z$ polar and azimuthal angles and so it is not straightforward to
construct the $\phi^\prime_-$ distribution.\s

In contrast, the $\theta_-$ distribution can be measured unambiguously.
Integrating the distribution over the lepton azimuthal angle $\phi_-$
casts the density matrix into a diagonal form
\begin{eqnarray}
  \rho^Z_{\lambda_Z,\lambda_Z}
= \frac{1}{4}\,{\rm diag}\left(1+c^2_- + 2A_\ell\, c_-,\ \
                          2s^2_-,\ \
                          1+c^2_- - 2A_\ell\, c_-\right) \quad
                          \mbox{with} \quad
                          \lambda_Z =+1,\, 0,\, -1\,,
\end{eqnarray}
depending on the reconstructible polar angle $\theta_-$. Furthermore,
integrating the correlated distribution over the azimuthal angle $\phi_Z$
also washes out the effects due to the off-diagonal components of
the $X_2$ polarization density matrix $\rho^{X_2}$ and leads to
the correlated polar-angle distribution given by
\begin{eqnarray}
  \frac{d\Gamma}{d\cos\theta_Z d\cos\theta_-}
&=& \frac{3 \kappa_Z\, {\rm B}(Z\to\ell^-\ell^+)}{64\pi m_2}
  \sum^{j_2}_{\lambda_2=-j_2}\,
  \rho^{X_2}_{\lambda_2,\lambda_2}\,
  \sum_{\lambda^\prime_Z}
  \sum_{\lambda_Z,\,\sigma_Z}\,
  \left[\, d^1_{\lambda^\prime_Z,\,\lambda_Z}(\omega_{_Z})
      d^1_{\lambda^\prime_Z,\, \sigma_Z}(\omega_{_Z})\,\right]\,
 \nonumber\\
&\times &
   \sum^{j_1}_{\sigma_1=-j_1}
   {\cal C}_{\lambda_Z,\sigma_1}\,
   {\cal C}^*_{\sigma_Z,\sigma_1}\,
   \left[\, d^{j_2}_{\lambda_2,\lambda_Z-\sigma_1}(\theta_Z)
       d^{j_2}_{\lambda_2,\sigma_Z-\sigma_1}(\theta_Z)\,\right]\,
   \rho^Z_{\lambda^\prime_Z,\lambda^\prime_Z}(\theta_-)\,,
\label{eq:polar_angle_correlations}
\end{eqnarray}
with the $X_2$ polarization density matrix $\rho^{X_2}$ depending on
the $X_2$ production mechanism and with the constraints $|\lambda_Z-\sigma_1|\leq j_2$
and $|\sigma_Z-\sigma_1|\leq j_2$ on the summation over the $X_1$
helicities as well as the $Z$ helicities $\pm 1$ and $0$.\s

\section{A Specific Example}
\label{sec:two_specific_examples}

As a concrete example of the correlated production-decay process
(\ref{eq:combined_eex2x1_x2zx1}),
we consider the production of a nondiagonal pair of two lighter
neutralinos $\tilde{\chi}^0_2$ and $\tilde{\chi}^0_1$ among the four
neutralinos, all of which are mixtures of U(1)$_Y$ and SU(2)$_L$ gauginos
$\tilde{B}$ and $\tilde{W}_3$ and two Higgsinos $\tilde{H}^0_1$ and
$\tilde{H}^0_2$ and  are spin-1/2 Majorana fermions in the MSSM.
In this example, we assume that the two-body decay $\tilde{\chi}^0_2\to Z
\tilde{\chi}^0$  is kinematically allowed, i.e. the second neutralino
mass is greater than the sum of the first neutralino mass and the
$Z$-boson mass~\cite{Choi:2003fs}. For notational convenience and
consistency, we set $\tilde{\chi}^0_2=X_2$ and $\tilde{\chi}^0_1=X_1$
in the following.\s

Generally, the production process $e^-e^+\to X_2 X_1$ has the contributions
from $t$- and $u$-channel selectron exchanges as well as a $s$-channel
$Z$ exchange. Nevertheless, for a simple demonstration without too
much loss of generality in the context of the present work, we assume
all the selectron-exchange contributions to be decoupled due to
sufficiently large selectron masses as in the context of the so-called split
supersymmetry scenario~\cite{Giudice:2004tc,ArkaniHamed:2004yi}, while
maintaining only the $s$-channel $Z$ contribution. In this case, for both
the production process $e^-e^+\to X_2 X_1$ and two-body decay $X_2 \to Z X_1$,
it is sufficient to consider in addition to the standard $Z\ell\ell$ vertices
the  $X_2X_1Z$ vertices whose expressions are
given in terms of a complex coupling by
\begin{eqnarray}
&&  \langle X_{2R} |Z| X_{1R}\rangle
  = \langle X_{1R} |Z| X_{2R}\rangle^*
  = +g_{_Z}\, {\cal Q}\,,
\label{eq:x2_x1_zr}\\
&& \langle X_{2L}\, |Z| X_{1L}\rangle
  = \langle X_{1L}\, |Z| X_{2L}\rangle^*
  = -g_{_Z}\, {\cal Q}^*\,,
\label{eq:x2_x1_zl}
\end{eqnarray}
for the right and left chiral modes with $g_{_Z}=e/c_W s_W$ and
the normalized coupling ${\cal Q}=(N_{13} N^*_{23}-N_{14} N^*_{24})/2$
in terms of the unitary $4\times 4$ matrix $N$ rotating the gauge eigenstate
basis to the mass eigenstate basis for diagonalizing the neutralino
mass matrix~\cite{Choi:2001ww}. Therefore, the axial-vector and vector
couplings are purely  real and purely imaginary, respectively.\s

The production transition amplitude for the process $e^-e^+\to X_2 X_1$ can be
expressed as a sum of two-current products as follows:
\begin{eqnarray}
  {\cal T}(e^-e^+\to X_2 X_1)
= g^2_Z D_Z(s)\, \sum_{a,b=\pm} Q_{ab}\,
  \left[\bar{v}(e^+)\gamma_\mu P_a u(e^-)\right]\,
  \left[\bar{u}(X_2)\gamma^\mu P_b v(X_2)\right]\,,
\label{ee_x2_x1_production_amplitude_example}
\end{eqnarray}
in terms of four bilinear charges, defined by the chiralities of the
associated electron and neutralino currents with
$P_\pm = (1+\pm\gamma_5)/2$. Explicitly, the normalized bilinear charges
are
\begin{eqnarray}
Q_{++} = c_+\, {\cal Q}^*\,, \quad
Q_{+-} = -c_+\, {\cal Q}\,, \quad
Q_{-+} = c_- \, {\cal Q}^*\,, \quad
Q_{--} = -c_- \, {\cal Q}\,,
\label{eq:normalized_bilinear_charges}
\end{eqnarray}
with the normalized $Z\ell\ell$ right- and left-chiral couplings
$c_+=s^2_W$ and $c_-=s^2_W-1/2$. Ignoring the electron mass, the electron
and positron helicities are opposite to each other in all amplitudes
so that the reduced production helicity amplitudes
$T_{\sigma,-\sigma; \lambda_2,\lambda_1} = g^2_Z\, s\, D_Z(s)\,
\langle \sigma; \lambda_2,\lambda_1 \rangle$ with
$\sigma,\lambda_2,\lambda_1=\pm 1/2=\pm $ are written in a compact
form as
\begin{eqnarray}
    \langle +;\pm,\pm\rangle
&=& \sqrt{2} c_+\,[\,\mu_+ \sqrt{1-\mu^2_-}\, \imath\, {\rm Im}({\cal Q})
             \pm \mu_- \sqrt{1-\mu^2_+}\, {\rm Re}({\cal Q})\,]\,,
\label{eq:production_amplitude_+pmpm} \\
    \langle -;\pm,\pm\rangle
&=& \sqrt{2} c_-\, [\,\mu_+ \sqrt{1-\mu^2_-}\, \imath\, {\rm Im}({\cal Q})
             \pm \mu_- \sqrt{1-\mu^2_+}\, {\rm Re}({\cal Q})\,]\,,
\label{eq:production_amplitude_-pmpm} \\
    \langle +;\pm,\mp\rangle
&=& 2 c_+\, [\,\sqrt{1-\mu^2_-}\, \imath\, {\rm Im}({\cal Q})
             \mp \sqrt{1-\mu^2_+}\, {\rm Re}({\cal Q})\,]\,,
\label{eq:production_amplitude_+pmmp}\\
    \langle -;\pm,\mp\rangle
&=& 2 c_-\, [\,\sqrt{1-\mu^2_-}\, \imath\, {\rm Im}({\cal Q})
             \mp \sqrt{1-\mu^2_+}\, {\rm Re}({\cal Q})\,]\,,
\label{eq:production_amplitude_-pmmp}
\end{eqnarray}
with the normalized dimensionless factors $\mu_\pm=(m_2\pm m_1)/\sqrt{s}$. We note that
CP is violated if both the real and imaginary parts of the complex factor
${\cal Q}$ are non-zero, as can be checked with the relation in
Eq.$\,$(\ref{eq:production_cp_relation}).
On the other hand, the Majorana condition
in Eq.$\,$(\ref{eq:production_majorana_condition}) is satisfied with $J=1$ and
the overall intrinsic parity of $\eta^P_{\rm M}=+1$. \s

The same complex factor ${\cal Q}$ appearing in Eq.$\,$(\ref{eq:normalized_bilinear_charges})
enables us to describe the two-body decay $X_2\to Z X_1$ fully. The reduced decay
helicity amplitudes in the $X_2$RF, which is independent of the $X_2$ helicity due to
angular momentum conservation, read
\begin{eqnarray}
    {\cal C}_{\pm\pm}
&=& \sqrt{2}\, g_{_Z}\,
    [\, \sqrt{m_-^2-m^2_Z}\,\, {\rm Im}({\cal Q})
    \mp \imath\, \sqrt{m_+^2-m^2_Z}\,\, {\rm Re}({\cal Q})\, ]\,,
\label{eq:reduced_decay_helicity_amplitudes_pmpm}\\
   {\cal C}_{0\,\pm}
&=& g_{_Z}\,
    \left[\, \frac{m_+}{m_Z} \sqrt{m_-^2-m^2_Z}\,\, {\rm Im}({\cal Q})
    \mp \imath\, \frac{m_-}{m_Z} \sqrt{m_+^2-m^2_Z}\,\, {\rm Re}({\cal Q})\, \right]\,,
\label{eq:reduced_decay_helicity_amplitudes_0pm}
\end{eqnarray}
for $\lambda_{Z}=\pm1, 0=\pm, 0$ and $\sigma_1=\pm 1/2=\pm$ with the convention
$m_\pm = m_2\pm m_1$ introduced for notational convenience.
The remaining reduced helicity amplitudes ${\cal C}_{\pm\mp}$ are vanishing
due to angular momentum conservation.
Furthermore, all the angular dependent parts are encoded solely in Wigner $d$
functions. The CP relation in Eq.$\,$(\ref{eq:x2_z_x1_cp_relation})
is violated again if the coupling ${\cal Q}$ is neither purely real nor purely
imaginary. Note that the Majorana condition (\ref{eq:x2_z_x1_majorana_condition})
is valid with the combined intrinsic parity of $\eta^C_{\rm M}=+1$.\s

Since the lightest neutralino escapes undetected and the heavier neutralino
decays into the invisible lightest neutralino and a $Z$ boson, the production
angle $\theta_2$ cannot be determined unambiguously for non-asymptotic
energies.\s

To describe the electron and positron polarizations in a general setting,
the reference frame must be fixed. The electron-momentum direction can be
used to define the $z$-axis. If the electron beam is transversely polarized,
the direction of transverse polarization is set to be the $x$-axis. In any
case, because the azimuthal angle $\phi_2$ of the $X_2$ momentum cannot be
reconstructed with the invisible $X_1$, we consider only the longitudinally
polarized electron and positron beams. Then, the polarized differential
production cross section is given in terms of the degrees of electron and
positron longitudinal polarizations, $P_L$ and $\bar{P}_L$, by
\begin{eqnarray}
  \frac{d\sigma}{d\cos\theta_2}
= \frac{\pi\alpha^2_Z\kappa_{21}}{4}\,s\,|D_Z(s)|^2\,
  \left[\,(1-P_L\bar{P}_L)\, \Sigma_{\rm unp}
         +(P_L-\bar{P}_L)\,  \Sigma_{LL}\right]\,,
\label{eq:polarized_production_distribution}
\end{eqnarray}
with $\kappa_{21}=\lambda^{1/2}(1,m^2_2/s,m^2_1/s)$ and $\alpha_Z=g^2_Z/4\pi$.
The coefficients $\sum_{\rm unp}$ and $\sum_{LL}$
depend on the polar angle $\theta_2$ and the $e^-e^+$ CM energy but not on the
azimuthal angle $\phi_2$ any more. Their expressions are given in terms of the
chiral complex factor ${\cal Q}$ by
\begin{eqnarray}
 \Sigma_{\rm unp}
&=& (c^2_+ + c^2_-)\,
    \left\{[1-(\mu^2_2-\mu^2_1)^2+\lambda_{21}\cos^2\theta_2]\, |{\cal Q}|^2
   -4 \mu_2\mu_1 {\rm Re}({\cal Q}^2)\right\}\,,
\label{eq:unpolarized_part} \\[1mm]
 \Sigma_{LL}
&=& (c^2_+ - c^2_-)\,
    \left\{[1-(\mu^2_2-\mu^2_1)^2+\lambda_{21}\cos^2\theta_2]\, |{\cal Q}|^2
   -4 \mu_2\mu_1 {\rm Re}({\cal Q}^2)\right\}\,,
\label{eq:longitudinal_part}
\end{eqnarray}
with $\mu_{1,2}=m_{1,2}/\sqrt{s}$. We note that the coefficients
$\Sigma_{\rm unp}$ and $\Sigma_{LL}$ are FB symmetric with respect to
the polar angle $\theta_2$, as guaranteed by the Majorana condition,
and as a matter of fact they are proportional to each other, rendering
the normalized  angular distribution independent of the beam polarizations.
In any case, the $e^-$ and $e^+$ beam polarizations can be employed
for increasing the production rate. \s

The chiral structure of the neutralinos can be also inferred from the
polarization of the neutralinos. The degree of longitudinal $X_2$
polarization for longitudinally polarized electron and positron beams
is given in a simple factorized form as
\begin{eqnarray}
   P_L (\theta_2)
= {\cal P}_{ee}(P_L,\bar{P}_L)\,
    \frac{\left[\,(1-\mu^2_2-\mu^2_1)\,|{\cal Q}|^2
              -2\mu_2\mu_1{\rm Re}({\cal Q}^2)\,\right]\,\cos\theta_2 }{
          [1-(\mu^2_2-\mu^2_1)^2+\kappa^2_{21}\cos^2\theta_2]|{\cal Q}|^2
            -4\mu_2\mu_1{\rm Re}({\cal Q}^2)}\,,
\label{eq:x2_longitudinal_polarization}
\end{eqnarray}
with the effective $e^-e^+$ longitudinal-polarization factor $P_{ee}$ given by
\begin{eqnarray}
  {\cal P}_{ee}(P_L,\bar{P}_L)
= \frac{(1-P_L\bar{P}_L) A_e + P_L-\bar{P}_L}{
         1-P_L\bar{P}_L+(P_L-\bar{P}_L) A_e}\,,
\label{eq:ee_longitudinal_polarization_factor}
\end{eqnarray}
with $A_e=(c^2_+ - c^2_-)/(c^2_+ + c^2_-)=2v_e a_e/(v^2_e+a^2_e)\simeq
-0.16$~\cite{Zyla:2020zbs}.
Consequently, the $e^-$ and $e^+$ longitudinal beam polarizations change 
the overall size of the production rate but they do not affect the angular 
distribution of the $X_2$ longitudinal polarization. 
Furthermore, the longitudinal polarization is
FB antisymmetric with respect to the polar angle $\theta_2$ so that the $X_2$
particle is unpolarized on average after integrating over the production
polar-angle $\theta_2$. \s

After the $\theta_2$ integration is taken, we obtain the normalized correlated
polar-angle distribution, which is independent of the production mechanism, as
\begin{eqnarray}
   \frac{1}{\Gamma} \frac{d\Gamma}{d\cos\theta_Z d\cos\theta_-}
= \frac{3}{4} \sum_{\lambda^\prime_Z}\,
    W_{\lambda^\prime_Z,\lambda^\prime_Z}(\omega_{_Z})\,
    \rho^Z_{\lambda^\prime_Z,\lambda^\prime_Z}(\theta_-)\,,
\label{eq:normalized_correlated_polar_angle_distributions}
\end{eqnarray}
where the so-called Wick distribution functions
$W_{\lambda^\prime_Z,\lambda^\prime_Z}(\omega_{_Z})$ 
are defined as~\cite{Choi:2019aig}
\begin{eqnarray}
 W_{\lambda^\prime_Z,\lambda^\prime_Z}(\omega_{_Z})
 =
 \sum_{\lambda_Z,\sigma_1}\,
    [\, d^1_{\lambda^\prime_Z,\lambda_Z}(\omega_{_Z})\,]^2\,
      |{\cal C}_{\lambda_Z,\sigma_1}|^2/
    \sum_{\lambda_Z,\sigma_1}\,
    |{\cal C}_{\lambda_Z,\sigma_1}|^2\quad
    \mbox{with}\quad
    |\lambda_Z-\sigma_1|\leq 1/2\,,
\label{eq:wick_distribution_function}
\end{eqnarray}
of which the sum is normalized to unity. The Majorana condition (\ref{eq:x2_z_x1_majorana_condition}) on the reduced decay helicity amplitudes
guarantees $W_{++}(\omega_{_Z})=W_{--}(\omega_{_Z})$ leading to the absence of
the parity-violating distribution linear in $\cos\theta_-$. Consequently, like
the production polar-angle distribution, the decay $\theta_-$ distribution is
forward-backward symmetric. Explicitly, the normalized two-dimensional
correlated polar-angle distribution independent of the
magnitude of the complex factor ${\cal Q}$ is given by
\begin{eqnarray}
 \frac{1}{\Gamma} \frac{d\Gamma}{d\cos\theta_Z d\cos\theta_-}
 =\frac{1}{4}
   \left[\, 1-\frac{1}{4}\, \eta_{\cal Q}\,
            (3\cos^2\omega_{_Z}-1)(3\cos^2\theta_- -1)\,\right]\,,
\label{eq:explicit_correlated_polar_angle_distribution}
\end{eqnarray}
where the $\alpha_{\cal Q}$-dependent coefficient $\eta_{\cal Q}$ is defined as

\begin{eqnarray}
 \eta_{\cal Q}
= \frac{1}{2}\,
  \frac{[(m_2+m_1)^2-m^2_Z]\,[(m_2-m_1)^2-m^2_Z]}{
        [(m_2+m_1)^2+2m^2_Z]\, [(m_2-m_1)^2-m^2_Z]+12m_2m_1m^2_Z \cos^2\alpha_{\cal Q}}\,,
\label{eq:q_dependent_coefficient}
\end{eqnarray}
in terms of the phase angle $\alpha_{\cal Q}$ of
${\cal Q}=|{\cal Q}|(\cos\alpha_{\cal Q}+\imath \sin\alpha_{\cal Q})$.
If the lepton polar-angle dependence is not taken into account,
the $\theta_Z$ distribution is simply isotropic, i.e. independent of the polar
angle $\theta_Z$. On the other hand, the $\theta_-$ dependence is sensitive
to the boost factor $\beta_2$ of the decaying particle $X_2$. For instance,
if the particle $X_2$ is at rest, the vanishing Wick helicity rotation angle
$\omega_{_Z}$ renders the $\theta_-$ distribution maximally dependent on
the coefficient $\eta_{\cal Q}$. \s

\vskip 0.3cm

\begin{figure}[tbh]
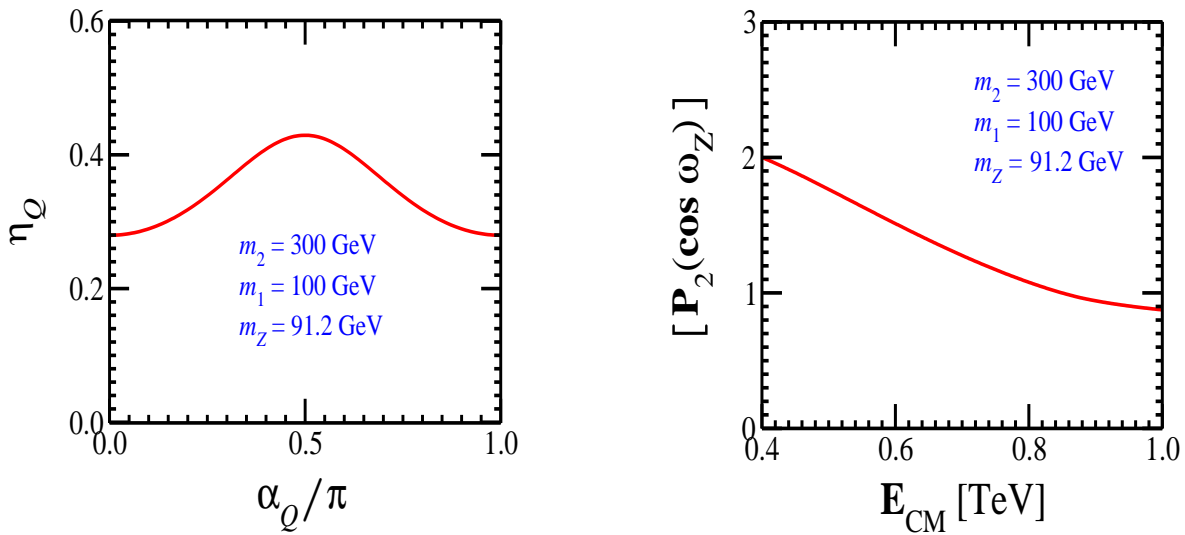

\begin{center}
\includegraphics[width=6.8cm, height=7.0cm]{eta_Q.eps}\hskip 2.0cm
\includegraphics[width=6.8cm, height=7.0cm]{p_2_integral.eps}
\caption{\it (Left) The coefficient $\eta_{\cal Q}$ is shown as a function of
             the phase $\alpha_{\cal Q}$. (Right) The integral
             $[ P_2(\cos\omega_{_Z}) ]$ of the second Legendre polynomial
             $P_2$ of $\cos\omega_{_Z}$  is shown as a function of the $e^-e^+$
             CM energy $\sqrt{s}=E_{\rm CM}$ (right). For this simple
             illustration, the range of  $\sqrt{s}=E_{\rm CM}$ is taken
             to be from $0.4\, {\rm TeV}$ to $1.0\, {\rm TeV}$.}
\label{fig:eta_q_p_2_integral}
\end{center}
\end{figure}

For an explicit numerical illustration, we set the following values for the
$X_2$ and $X_1$ masses
\begin{eqnarray}
m_2=300\, {\rm GeV} \quad\mbox{and}\quad  m_1=100\, {\rm GeV}\,,
\label{eq:chosen_mass_values}
\end{eqnarray}
while varying the $e^-e^+$ CM energy $\sqrt{s}=E_{\rm CM}$ from 0.4 TeV to 1.0 TeV.
The left side of Figure~\ref{fig:eta_q_p_2_integral} shows the dependence of the
coefficient $\eta_{\cal Q}$ on the phase $\alpha_{\cal Q}$.
By measuring the coefficient we can determine the phase $\alpha_{\cal Q}$ up to
a two-fold discrete ambiguity. Unless  $\alpha_{\cal Q}$ is $0, \pi/2$ or $\pi$,
i.e. unless ${\cal Q}$ is purely real or imaginary, CP is violated in the neutralino
system.  The right side of Figure~\ref{fig:eta_q_p_2_integral} shows
the integral of $P_2(\cos\omega_{_Z})=(3\cos^2\omega_{_Z}-1)/2$ over the polar
angle $\theta_Z$ as a function of the $e^-e^+$ CM energy $\sqrt{s}=E_{\rm CM}$.
For a simple illustration the $E_{\rm CM}$ range is taken from 0.4 TeV (identical to
the threshold energy of $m_2+m_1$) to 1.0 TeV. The integral value is monotonically 
decreasing implying that the sensitivity to the complex factor ${\cal Q}$ is 
maximal at the production threshold. In contrast, the production cross section is 
increasing with the CM energy near the threshold. So, there exists a specific 
value of $E_{\rm CM}$ above the threshold for the optimal sensitivity to ${\cal Q}$. \s

\section{Conclusions}
\label{sec:conclusions}

In this paper we have made a general and systematic model-independent study 
of correlated distributions connected to the production process 
$e^-e^+\to X_2 X_1$ of two Majorana particles $X_2$ and $X_1$ with
different masses and arbitrary spins, followed by two sequential 2-body
decays, $X_2\to Z X_1$  and $Z\to\ell^-\ell^+$ with $\ell=e$ or $\mu$, with
invisible $X_1$. The constraints due to CP invariance and the Majorana
condition were discussed. Formally, a proper Wick helicity rotation and
an azimuthal-angle adjustment were taken into account for combining the
production and decay helicity amplitudes derived in the most compact form 
from an analytic point of view and for describing a few general properties 
of the combined production-decay process involving two Majorana particles
in a transparent way. Then, a specific example with the non-diagonal 
pair of two lighter neutralinos has been investigated for demonstrating 
the validity of all the worked-out general properties in a concrete and 
detailed manner.\s

In relation to this work we are at present analyzing the general structure
of the $ZX_2X_1$ interaction vertices of a $Z$ boson and two Majorana
particles $X_2$ and $X_1$ with any integer and/or half-integer spins and,
furthermore, we plan to probe that of the interaction vertices of three 
Majorana particles with arbitrary spin combinations. This research project, 
of which the outcome will be reported soon elsewhere, is a natural extension 
of several previous works~\cite{Kayser:1982br,Kayser:1984ge,Boudjema:1988zs,
Boudjema:1990st,Nieves:1996ff,Nieves:2013csa}.\s

\section*{Acknowledgment}
We thank Ji Ho Song for his early-stage contributions to the present work.
The work was in part by the Basic Science Research Program of Ministry of
Education through National Research Foundation of Korea (Grant No.
NRF-2016R1D1A3B01010529) and in part by the CERN-Korea theory collaboration.

\end{document}